\pdfoutput=1
\documentclass[useAMS,usenatbib,usegraphicx]{mn2e}
\usepackage{color}
\usepackage{aas_macros}

\usepackage{hyperref}
\usepackage{breakurl}

\newcommand{\bdm}{\begin{displaymath}}
\newcommand{\edm}{\end{displaymath}}
\newcommand{\be}{\begin{equation}}
\newcommand{\ee}{\end{equation}}
\newcommand{\bea}{\begin{eqnarray}}
\newcommand{\eea}{\end{eqnarray}}
\newcommand{\hmpc}{\ensuremath{h^{-1}{\rm Mpc}}}

\title[SMAD Modeling of Galaxy Statistics]{Connecting Galaxies with Halos Across Cosmic Time: Stellar mass assembly distribution modeling of galaxy statistics}

\author[M.~R.~Becker]{M.~R.~Becker$^{1,2}$ \\
$^{1}${KIPAC, Physics Department, Stanford University, Stanford, CA 94305}. \\
$^{2}${KIPAC, SLAC National Accelerator Laboratory, Menlo Park, CA 94025.}\\
}

\begin{document}

\maketitle

\label{firstpage}

\begin{abstract}
In this work, I explore an empirically motivated model for investigating the relationship between galaxy stellar masses, 
star formation rates and their halo masses and mass accretion histories. The core statistical quantity in this model is the stellar mass 
assembly distribution, $P(dM_{*}/dt|\mathbf{X},a)$, which specifies the probability density distribution of stellar mass assembly 
rates given a set of halo properties $\mathbf{X}$ and epoch $a$. Predictions from this model are obtained by integrating the stellar mass 
assembly distribution (SMAD) over halo merger trees, easily obtained from modern, high-resolution $N$-body simulations. 
Further properties of the galaxies hosted by the halos can be obtained by post-processing the stellar mass assembly histories with 
stellar population synthesis models. In my particular example implementation of this model, I use the \citet{behroozi13a} constraint 
on the median stellar mass assembly rates of halos as a function of their mass and redshift to construct an example parameterization 
of $P(dM_{*}/dt|\mathbf{X},a)$. This SMAD is then integrated over individual halo mass accretion histories from $N$-body merger 
trees starting at z = 4, using simple rules to account for merging halos.  I find that this a simple model can reproduce qualitatively 
the bimodal features of the low-redshift galaxy population, including the qualitative split in the two-point clustering as a function of specific 
star formation rate. These results indicate that models which directly couple halo and galaxy growth through simple efficiency functions 
can naturally predict the star formation rate bimodality in higher-order statistics of the galaxy field, such as its two-point correlations or 
galactic conformity signals. 
\end{abstract}

\begin{keywords}
\end{keywords}

\section{Introduction}
The connection between the growth of galaxies and dark matter halos in the cosmological context of a $\Lambda$CDM 
universe is one of the primary concerns of galaxy formation theory. In the modern era, a variety of approaches have been taken to study 
this important problem. These range from \textit{ab initio} computations of the formation of galaxies in large, supercomputer 
simulations (see, e.g., \citeauthor{somerville2014}~\citeyear{somerville2014} for a review) to fitting empirical models to observations 
of galaxies in order to infer the properties of their host dark matter halos \citep[i.e., the halo occupation distribution, HOD, or conditional 
luminosity function, CLF, models; see, e.g.,][]{berlind2002,vdb2007,zehavi2011,cacciato2013}. Between these two extremes, other approaches 
combine directly high-resolution, dark-matter-only 
supercomputer simulations of structure formation with simplified prescriptions for galaxy formation. These approaches include 
semi-analytic models (SAMs) where one attempts to parameterize directly the physics of galaxy formation \citep[e.g.,][see \citeauthor{baugh2006}~\citeyear{baugh2006} or 
 \citeauthor{somerville2014}~\citeyear{somerville2014} for a review]{white1991,cole1994,somerville1999}, or various forms of abundance matching, like subhalo abundance matching 
 \citep[SHAM, e.g.,][]{kravtsov2004,conroy2006,conroy2009,reddick2013} and 
 conditional abundance matching \citep[CAM,][]{hearin2013,hearin2014,watson2015}, that take the empirically \textit{anzatz} 
 that the most massive galaxies occupy the largest halos on average. Yet another approach along these lines combines observations at multiple 
epochs with flexible models of the stellar mass-to-halo mass relation in order to constrain the growth of stellar mass in halos across cosmic time 
\citep[e.g.,][]{conroy2009,moster2010,behroozi13a}.

None of these approaches alone is currently sufficient to fully describe the wealth of observations of galaxies. A key advantage of models 
like HOD/CLFs, SHAM/CAMs or models similar to that in \citet{behroozi13a}, is that they are formulated empirically, allowing for data to directly 
constrain how galaxies occupy halos, and when and where dark matter halos form galaxies. Alternatively, \textit{ab initio} simulations of 
galaxy formation offer the opportunity to directly test models of the primary physical processes that govern galaxy formation. SAMs can offer 
these insights as well. 

In this preliminary work, I explore a statistical approach to modeling the galaxy-halo connection across cosmic time through the 
stellar mass assembly distribution (SMAD)
\be
P(dM_{*}/dt|\mathbf{X},a)\ .
\ee
This probability density function describes the distribution of stellar mass assembly rates given a set of halo properties 
$\mathbf{X}$ and epoch $a$. It is important to note, as described further below, that many previous authors have built models 
very similar to that proposed in this work \citep[e.g.,][]{wang2007,mutch2013,lilly2013,lu2014,lu2015}. In this work, I propose to 
assume a form for the SMAD and then directly integrate it over the ensemble of halo mass accretion histories from high-resolution 
$N$-body simulations. 

For simplicity and practicality, I assume the \citet{behroozi13a} relationship between halo mass and stellar mass assembly rate 
without any modifications. This relationship is used to a build an example form of the SMAD, which is then applied directly 
to individual halo mass accretion histories from a high-resolution $N$-body simulation. Note that the \citet{behroozi13a} relation was 
originally constrained for median median mass accretion histories as opposed to individual mass accretion histories. 
I then perform some preliminary comparisons to local data, mostly from the Sloan Digital Sky Survey  \citep[SDSS,][]{sdss,sdss7}. 
I find that while my model does not perfectly reproduce statistics from local data, it captures many of the qualitative features of local data, 
including the specific star formation rate (sSFR) bimodality in the clustering of galaxies. This result, in agreement with the physical 
interpretation of galactic conformity through the effects of tidal forces on the mass accretion rate of dark matter halos \citep{hearin2015}, 
indicates that modeling galaxy stellar mass assembly in a way which is directly coupled to halo mass assembly can naturally predict 
the bimodal higher-order statistics of the galaxy density field. Future work exploring general parameterizations of the SMAD, directly fit 
to observational data, will improve the results presented here, but is well beyond the scope of this work. 

This work is organized as follows. In Section~\ref{sec:data}, I describe the simulation and observational data used in this work. 
In Section~\ref{sec:smad}, I further discuss the general principle of SMAD modeling and describe my example model. 
Section~\ref{sec:comp} compares the results of my example SMAD to the data from the local Universe, mostly the SDSS. Finally, 
I conclude in Section~\ref{sec:conc}.

\section{Simulation and SDSS Data}\label{sec:data}
In this work, I use a 250 \hmpc\ box with $2048^{3}$ particles, \texttt{c250}. The \texttt{c250} box has a flat $\Lambda$CDM cosmology with 
$\Omega_{m},\Omega_{b},\sigma_{8},n_{s},h = \{0.286,0.047,0.82,0.96,0.7\}$ with $h\equiv H_{0}/100\,{\rm kms^{-1}Mpc^{-1}}$. 
The box was run with \verb+LGadget-2+, an $N$-body only version of the \verb+Gadget-2+ code \citet{springel2005}. The initial conditions were generated 
with \verb+CAMB+ \citep{lewis2002} and second-order Lagrangian perturbation theory at $z=99$ with the \verb+2LPTic+ \citep{crocce2006}. 
One hundred snapshots were saved between $z\approx12$ and $z=0$, evenly spaced in the logarithm of the scale factor. Halos and merger 
trees were constructed with the \verb+ROCKSTAR+ \citep{rockstar} and \verb+CONSISTENT TREES+ codes \citep{ctrees}. All halos with at least 20 
particles are used in integrating the stellar mass assembly rates over the merger trees. However, when comparing to 
the low redshift SDSS data, only halos with maximum circular velocity greater than 55 ${\rm kms^{-1}}$ are used. 

I use the measurements of the projected correlation function of SDSS galaxies from \citet{watson2015}. These measurements were performed 
on the SDSS DR7 catalog \citep{sdss,sdss7} using all galaxies with stellar masses satisfying $\log_{10}(M_{*}/M_{\odot}) > 9.8$. The galaxies 
were split according to their star formation rates, computed using a combination of emission line indicators \citep[i.e., H$\alpha$][]{brinchmann2004,salim2007} 
and Dn4000 \citep{kauffmann2003} in the cases of no emission lines or AGN contamination. The catalog of star formation rates is publicly 
available.\footnote{\url{http://www.mpa-garching.mpg.de/SDSS/DR7.}} Finally, for some comparisons to data, I use directly the SDSS mock constructed by 
\citet{watson2015}. This mock catalog successfully predicts a large number of measurements from the SDSS, include the projected correlation function 
as a function of sSFR, galaxy-galaxy lensing as a function of sSFR, the colors and sSFR's of galaxies in galaxy groups, and the colors and sSFR's of 
central galaxies \citep{hearin2013,hearin2014,watson2015}.

\section{Connecting Galaxies with Halos Across Cosmic Time}\label{sec:smad}
In this work, I explore how one can use draws of merger trees from $N$-body simulations to perform the integral of the SMAD over the population of halos 
in $\Lambda$CDM-like universes.\footnote{Analytical integrals of this PDF over stochastic processes which model the mass accretion 
histories of halos may be a useful avenue for future work, but are neglected in this work.} This approach has several important features. 
\begin{enumerate}
\item The SMAD is statistical in nature, aiming to predict (or constrain!) the ensemble of stellar mass assembly rates (and thus galaxy properties) 
from the ensemble of halo properties. Note that any purely deterministic model is the limit $P(dM_{*}/dt|\mathbf{X},a) = \delta(dM_{*}/dt - f(\mathbf{X},a))$ 
where $f(\mathbf{X},a)$ is some function specifying the stellar mass assembly rate. 
\item The statistical nature of the SMAD means that one does not have to specify all (or any, see the next feature) of the physical processes which control 
star formation for any individual halo (and information about which may not be present in a dark matter-only simulation). Instead, the SMAD posits that 
on average halos of a given set of properties form stars at a given rate and that the scatter about this average is a property of the ensemble of halos, 
characterizing something intrinsic to galaxy formation itself. 
\item One can take a largely physics-agnostic approach by fitting $P(dM_{*}/dt|\mathbf{X},a)$ to observational data in the spirit of works which combine 
abundance matching or direct models of the stellar mass-to-halo mass relation across epochs \citep[e.g.,][]{conroy2009,moster2010,behroozi13a}, 
SHAM/CAM models \citep[e.g.,][]{kravtsov2004,conroy2006,hearin2013,hearin2014,watson2015} and HOD/CLF models \citep[e.g.,][]{berlind2002,zehavi2011,cacciato2013}.
\item Unlike other common statistical and empirical descriptions of the galaxy population, the SMAD explicitly connects galaxies across cosmic time for individual halos. In particular, 
integrating the SMAD over the merger history of dark matter halos assigns a star formation history to every halo in the simulation 
in an empirical manner. Thus it is possible to directly predict the colors/SEDs of galaxies, the rate of SNe as a function of epoch, the 
overall star formation rate density of the Universe as a function of epoch, etc. Thus SMAD modeling will be a key tool for connecting observations of 
galaxies across cosmic time.
\end{enumerate}
Note that this approach encapsulates that taken by many other authors \citep{wang2007,mutch2013,lilly2013,lu2014,lu2015} where deterministic, 
empirical parameterizations of the growth of galaxies are used with merger trees to predict galaxy properties.

Furthermore, the SMAD model is similar to SAMs, except for its largely empirical nature. In fact, SMAD models are properly thought of as hybrids 
between HOD/CLF or other empirical models and SAMs, retaining the empirical and statistical nature of these models while explicitly connecting halos across cosmic 
time like a SAM. SMAD models also have important relations to SHAM/CAM models. In particular, below I assume 
that $dM_{*}/dt\propto dM_{\rm vir, cen}/dt$. Thus any assembly bias \citep[e.g.][]{sheth2004,gao2005,wechsler2006,harker2006,bett2007} 
contained in mass accretion histories of dark matter halos will be 
captured in the star formation histories, star formation rates and stellar masses of halos. To the extent that SHAM/CAM models prefer proxies which 
themselves capture assembly bias, like $v_{max}$ or $v_{peak}$, one expects qualitatively similar predictions from the SMAD at any epoch to these models. This 
expectation is borne out below. Thus SMAD models can naturally incorporate assembly bias and could even constrain the degree to which stochasticity 
in the process of galaxy formation could erase assembly bias signatures present in the halo population.

\subsection{An Example SMAD Model}
The specific SMAD model in this work combines the results of \citet{behroozi13a} with the mass assembly history of halos from high-resolution $N$-body 
simulations to predict $dM_{*}(\mathbf{X},a)/dt$ and the full stellar mass assembly histories of the halos. Schematically, at the initial epoch stellar mass is 
seeded into the halos using a fiducial stellar mass-to-halo mass relation. Then the amount of stellar mass assembled by each halo is drawn 
from the SMAD given its properties $\mathbf{X}$ and the epoch $a$. Next, a simple set of rules is used to compute the stellar mass of the descendent 
halos from the progenitors in the tree. Finally, this process is repeated until the final epoch is reached. 

\begin{figure*}
\includegraphics[width=\columnwidth]{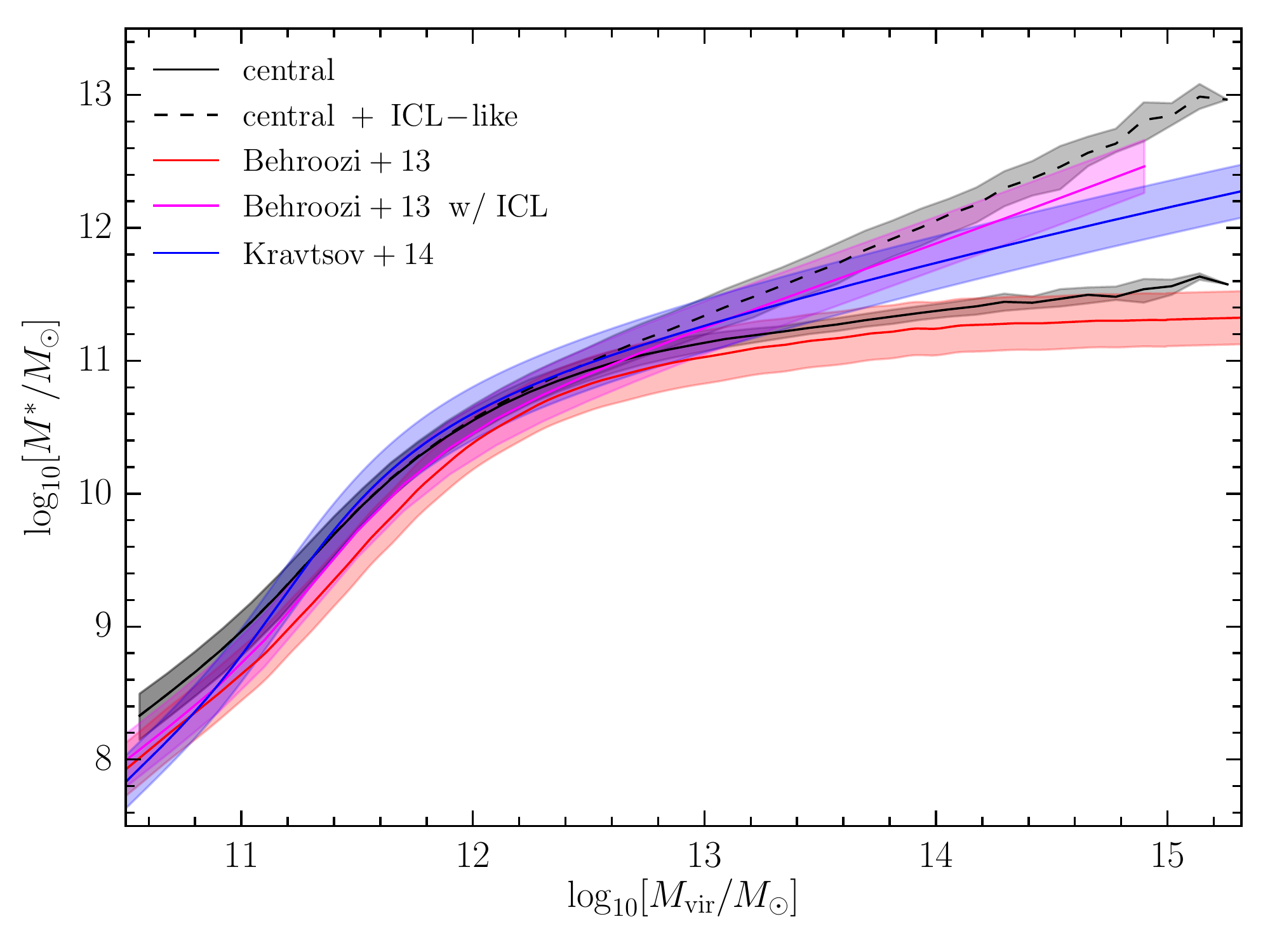}
\includegraphics[width=\columnwidth]{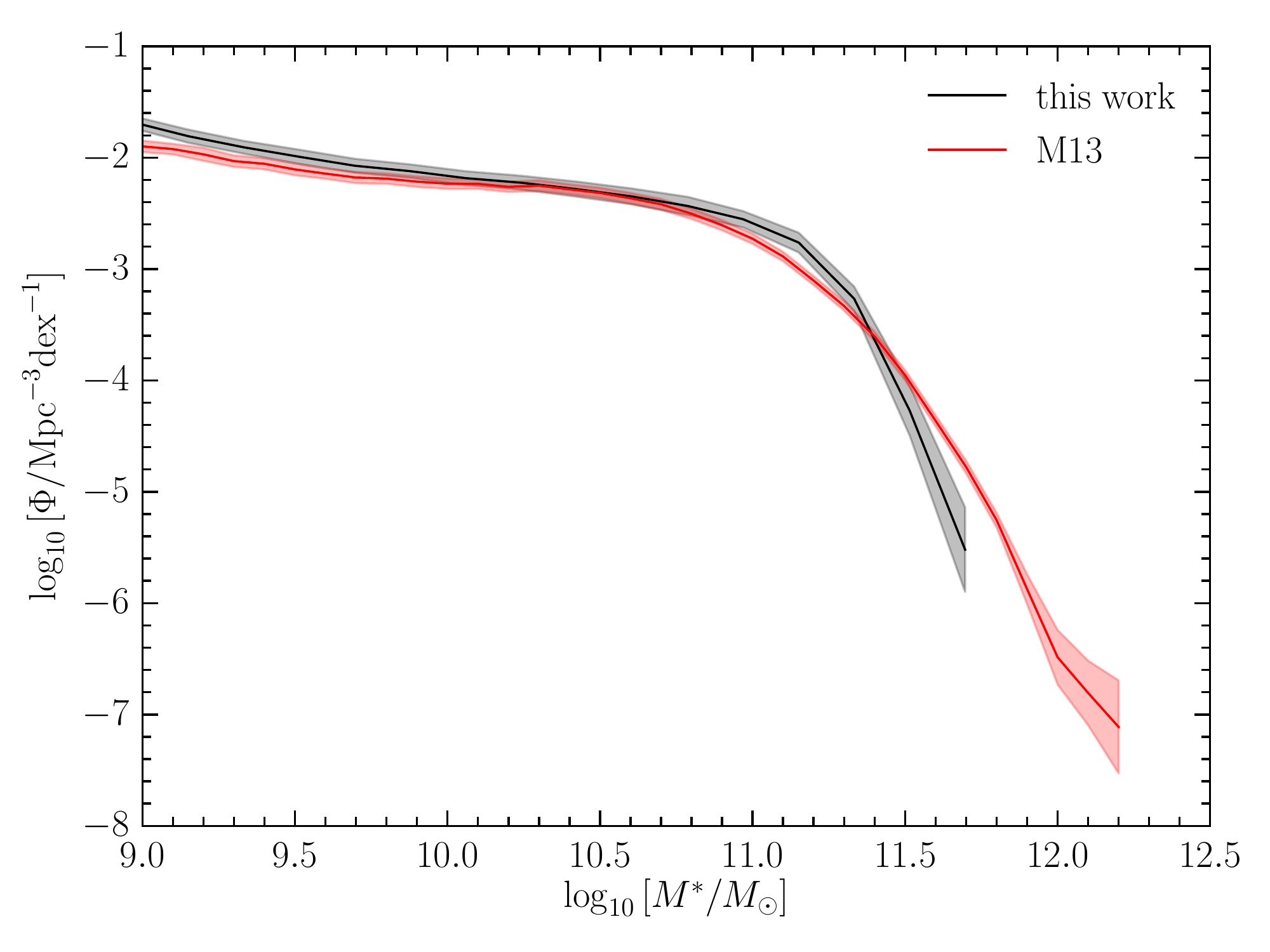}
\caption{The stellar mass-to-halo mass relation (left) and stellar mass function (right) of galaxies. In the left panel, the solid line shows the stellar 
mass-to-halo mass relation for the central galaxy in each halo, the dashed line shows the stellar mass in the central plus ICL-like variable for 
each halo, and the other lines show various constraints from low-redshift, mostly SDSS, data. The right panel shows the stellar mass function from this 
work and that from \citet{moustakas2013}.
\label{fig:smhm}}
\end{figure*}

\subsubsection{Assembling Stellar Mass with the SMAD}
The SMAD model I use in this work is given by drawing $dM_{*}/dt$ from 
\be\label{eqn:dms}
dM_{*}/dt \sim {\rm Logn}(10^{-12} {\rm yr^{-1}}\, M_{*},\epsilon) + \left.dM_{*}/dt\right|_{\rm B13}
\ee
where
\bea
\lefteqn{\left.dM_{*}/dt\right|_{\rm B13} = \left. dM_{*}(M_{\rm vir,peak},a)/dM_{\rm vir}\right|_{\rm B13}}&&\nonumber\\
&&\ \ \ \ \ \ \ \ \ \ \ \times \left\{
\begin{array}{ll}
\left.dM_{\rm vir,cen}/dt\right|_{i} & \left.dM_{\rm vir,cen}/dt\right|_{i} > 0\nonumber\\
0 & \left.dM_{\rm vir,cen}/dt\right|_{i} \le 0\nonumber\\
\end{array}\right.\ ,\nonumber
\eea
$\epsilon=0.25$, and ${\rm Logn}(\mu,\sigma)$ is a log-normal distribution of mean $\log_{10}\mu$ and scatter $\sigma$. I use base-10 logarithms in this work. 
$\left. dM_{*}(M_{\rm vir,peak},a)/dM_{\rm vir}\right|_{\rm B13}$ is the \citet{behroozi13a} result and $\left.dM_{\rm vir,cen}/dt\right|_{i}$ is the mass accretion history 
of halo $i$ in the simulation, as computed below. $M_{\rm vir,peak}(a)$ is the peak mass along the most massive progenitor branch of the merger tree of each halo and is the quantity used in 
\citet{behroozi13a} in order to define their parameterization of stellar mass assembly rates. The first term in this model is a rough, 
\textit{ad hoc} way to account for residual star formation in quenched galaxies, information about which is not present in dark-matter-only $N$-body simulations. 
Note that because of this choice, the quenched peak in the distribution of sSFR's in the mock is placed at the correct location and given roughly the 
correct width ``by hand.'' Furthermore, this model assumes that no dark matter stripping ever corresponds 
to the stripping of stellar mass. Such an assumption may not be warranted, but I leave exploring this issue to future work. Finally, note that the pseudo-evolution of $M_{\rm vir}$ 
\citep[i.e., changes in $M_{\rm vir}$ due to evolution in the over density used to define the halo as opposed to accretion of the mass onto the halo itself, see][]{diemer2013} 
cancels in the computation of the stellar mass assembly rates.

The mass accretion rate $dM_{\rm vir,cen}/dt$ is needed in order to compute $dM_{*}/dt$ above. For a halo that is not merging with another at a given epoch, 
\be
dM_{\rm vir,cen}/dt = \Delta M_{\rm vir,cen}/\Delta t \equiv \Delta M_{\rm vir}/\Delta t\ .
\ee
where $\Delta M_{\rm vir}$ is the mass difference between two epochs along the most massive progenitor branch of the merger tree and $\Delta t$ is 
the time spacing of the epochs. 

In a merger, the growth in the mass of the halo arises from smooth accretion and the merging of the halos themselves. In this method, the halos 
which merge have themselves already been forming stars, so that if one counted all of the mass change, one would effectively double count 
the mass growth and thus stellar growth. Thus during mergers one should adjust the mass growth using a definition similar to the following
\be\label{eqn:merger}
\Delta M_{\rm vir,cen} = \Delta M_{\rm vir} - \frac{M_{\rm vir,rest}}{M_{\rm vir,mmp} + M_{\rm vir,rest}}M_{\rm vir,desc}
\ee
where $M_{\rm vir,mmp}$ is the most massive progenitor halo mass and $M_{\rm vir,rest}$ is the sum of the masses of all other progenitors. 
Note that $\Delta M_{\rm vir}=M_{\rm vir,desc} - M_{\rm vir,mmp}$ so that this formula simply removes some fraction of the descendent halo's mass 
that is presumed to be due to the merging of halos besides the most massive progenitor. This fraction is fixed to the fraction of total mass before the merger. 
Halo mergers are not additive \citep[see, e.g.,][]{kazantzidis2006}, but as long as they are equally non-additive for all halos involved, this formula is roughly correct. 

However, in this work, I use the \citet{behroozi13a} constraints which correct for the effect of mergers on the stellar mass assembly, 
but are phrased in terms of the median halo mass assembly along the most massive progenitor track in the merger tree. These mass 
accretion rates include the merging of smaller structures. Thus in this work the correction due to the second term of equation~\ref{eqn:merger} is neglected. 

\begin{figure*}
\includegraphics[width=\textwidth]{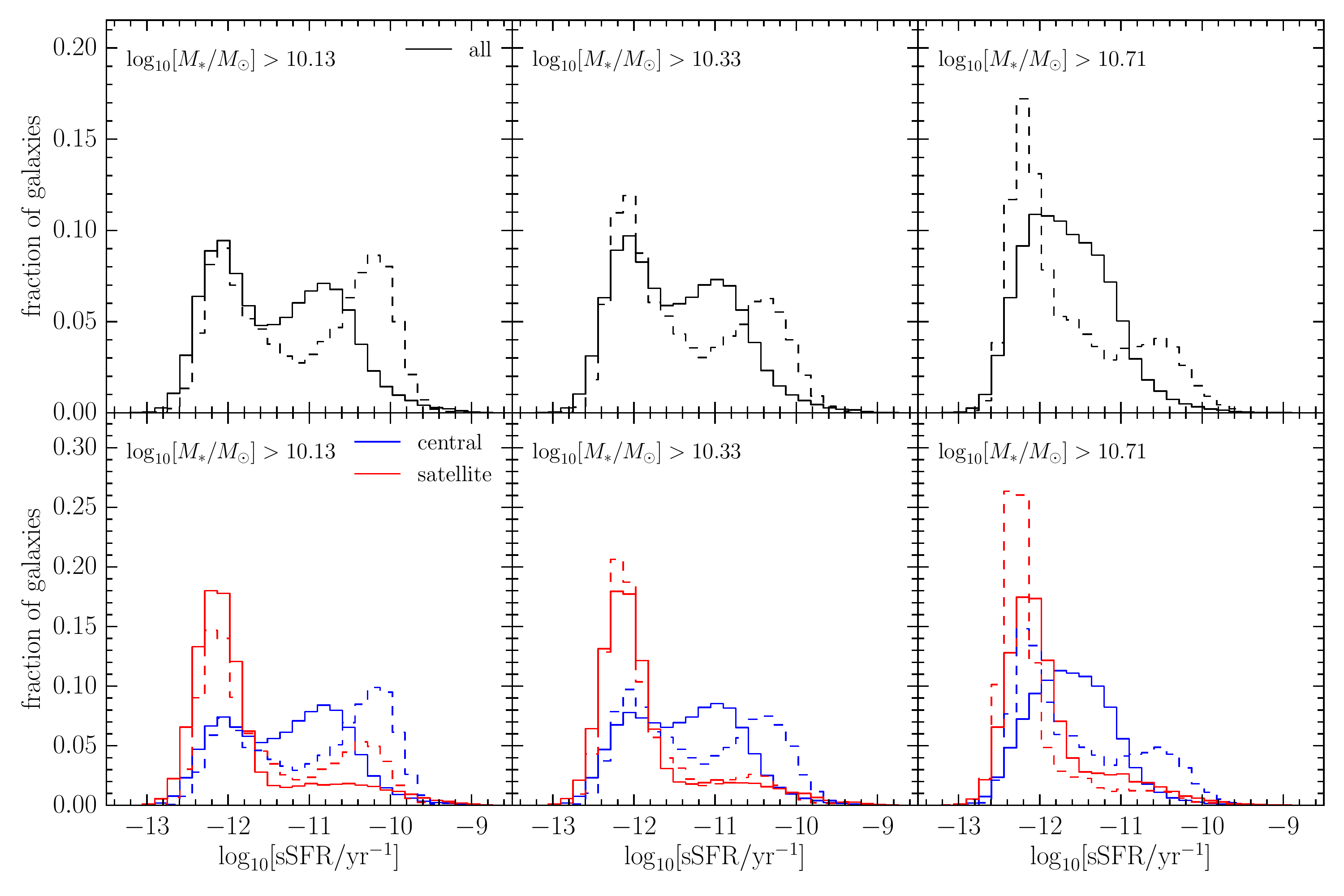}
\caption{The distribution of specific star formation rates as a function of stellar mass. The dashed lines in the top panel show the 
results from the SDSS as represented by the CAM mock catalog of \citet{watson2015}. The bottom panels show results split into 
central (\textit{blue}) and satellite ({\it red}) galaxies. The dashed lines in the bottom panel show the CAM prediction for centrals and satellites 
from \citet{watson2015}. \label{fig:ssfr}}
\end{figure*}

\subsubsection{Integrating the SMAD in Time}\label{sec:merger}
Once $dM_{*}/dt$ is determined for each progenitor halo, the total stellar mass formed is simply $dM_{*}/dt\times\Delta t$. The 
next task is then to assign the stellar mass to the halos and evolve it along the tree. In this work, I track two different stellar mass 
variables. The first is the stellar mass associated with the central galaxy of each halo or subhalo. The second is an intra-cluster light-like (ICL-like) variable 
that holds stellar mass deposited by halos not along the most massive progenitor branch during mergers between three or more halos at 
any epoch. 

Stellar mass is passed from progenitors to descendants according to the following rules:
\begin{description}
\item[\bf No Mergers:] For a halo that does not undergo a merger, the central stellar mass of the descendent is equal to the sum of the formed stellar mass
plus all of the previously formed central stellar mass. The stellar mass in the ICL-like variable is simply passed to the descendent. 
\item[\bf Binary Mergers:] For mergers of just two halos, the central stellar mass of both halos is added together into the central stellar mass of the final halo, along 
with any formed stellar mass. The ICL-like variables from both halos are added together and then passed to the descendent. 
\item[\bf $N>2$ Mergers:]  For halos which have more than two progenitors, only the central stellar mass of the most massive progenitor halo is passed to the central 
stellar mass of the descendent halo, along with any formed stellar mass. The central and ICL-like stellar masses of the other progenitors are added to the ICL-like 
stellar mass of the descendent halo.
\end{description}
Note that in all cases, the newly formed stellar mass is always added to the central variable of the descendent halo.

\subsubsection{Initial Conditions, Parameterization and Stellar Mass Loss}
With the rules and definitions above, a set of halos at any epoch can be evolved to another epoch given the merger tree. In order to 
seed halos with stellar mass at the initial epoch, redshift four in this work, I use the \citet{behroozi13a} constraint on the 
$M_{*}-M_{\rm vir,peak}$ relation at $z=4$. This choice is somewhat arbitrary and is motivated by the decreasing quantity galaxy 
population data above this epoch. Note that I make no adjustments to the \citet{behroozi13a} 
constraints, besides the minimal set of additional rules required to track the growth of stellar mass in halos in the central galaxy, 
in the ICL-like variable and the small amount of residual star formation seen in real galaxies. In particular, the details of the treatment of 
galaxy merging and the generation of intra-cluster light differ between \citet{behroozi13a} and this work. Thus there is no 
reason to expect exact quantitative agreement between the results of the two works. Finally, the stellar mass computed by this 
algorithm is the raw stellar mass assembled in the halos' histories. In order to account for stellar mass loss, I use the \citet{behroozi13a} fit to single stellar 
population evolutionary tracks in order correct the stellar masses.

\section{Results}\label{sec:comp}
In this section, I explore the results of my simple model, illustrating its one and two-point statistics. I focus explicitly on local, low-redshift data 
from the SDSS. Note that while the model agrees qualitatively with data from the SDSS, the quantitative agreement is in some cases lacking. 
Future work will explore how such models can be directly fit to SDSS and higher redshift data. Before comparing to the data, note that because 
this model is not tuned to match the SDSS stellar mass function when integrated over the merger trees, there is an overall disagreement between 
the stellar masses in the SDSS and those from the model (see Figure~\ref{fig:smhm}). Thus to compare the distributions, I compare galaxies of the same number density as 
those in the SDSS. For the thresholds of $\log_{10}[M_{*}/M_{\odot}]=\{9.8,10.2,10.6\}$ in the SDSS, I find that thresholds in my model of 
$\log_{10}[M_{*}/M_{\odot}]=\{10.13,10.33,10.71\}$ match the number density in the SDSS. For each stellar mass threshold, the split in 
specific star formation rate in the mocks is determined by requiring the same fraction of galaxies to be either star forming or passive as 
determined by a threshold of $\log_{10}[{\rm sSFR}/{\rm yr^{-1}}]=-11$ in the SDSS.

\subsection{One-point Galaxy Statistics}
Figure~\ref{fig:smhm} shows the stellar mass-to-halo mass relation and the stellar mass function. The right panel of Figure~\ref{fig:smhm} shows 
the stellar mass function from this work and that from \citet{moustakas2013}. The model error bars are from jackknife resampling of the simulation 
volume. While the agreement between the two is not perfect, the qualitative features, like the strongly decreased number 
density of massive galaxies, are correct. In the left panel of this figure, I show the results 
for the stellar mass of the central component (solid lines) and the stellar mass of the central plus the ICL-like component (dashed lines). 
I have also plotted the abundance matching relations from \citet{behroozi13a} and \citet{kravtsov2014}. I also show the \citet{behroozi13a} 
relation with the estimated ICL component included. The colored bands show the amount of scatter in abundance matching of $\approx0.2$~dex 
\citep[see, e.g.,][]{reddick2013}. Note that the \citet{behroozi13a} and \citet{kravtsov2014} relations differ in how the extended wings of light around BCGs are treated. 
The \citet{kravtsov2014} relation is expected to include more of the light around BCGs and yields higher stellar masses for the most massive halos. 
Matching the aperture used to estimate the stellar mass in the data with the effective aperture implied by the central stellar mass variable tracked 
above is difficult and beyond the scope of this work. However, my model does track the total stellar mass in both the central and ICL-like component, 
as inferred by \citet{behroozi13a}, reasonably well. Finally, note that my predictions also produce a scatter 
in the stellar mass-to-halo mass relation, as show by the grey band, that is smaller than, but close to the inferred scatter of $\approx$0.2 dex 
from abundance matching.

\begin{figure}
\includegraphics[width=\columnwidth]{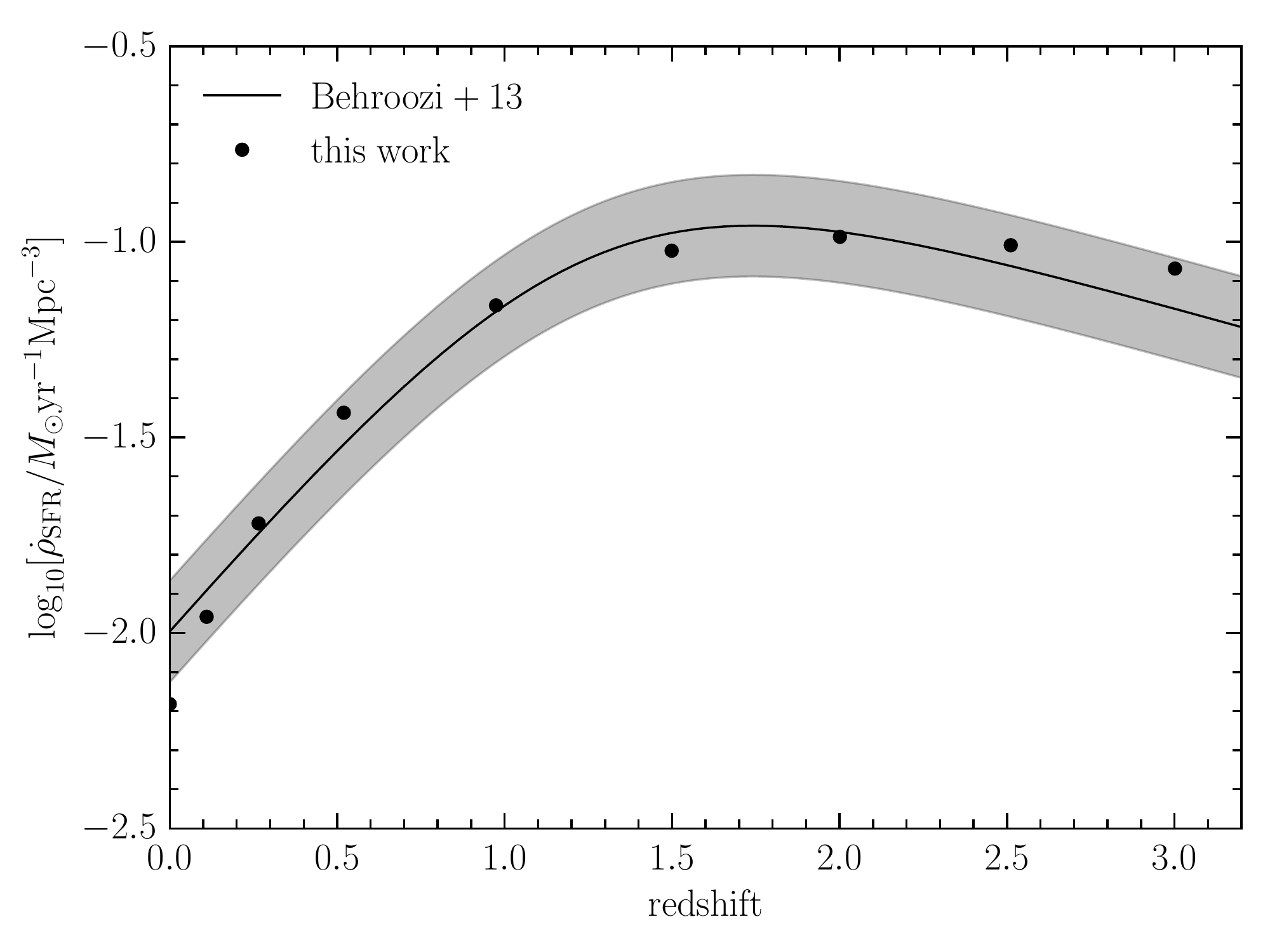}
\caption{The star formation rate density, $\dot\rho_{\rm SFR}$ as a function redshift. The points show the results of this work. The 
line with grey band shows the constraints from data on $\dot\rho_{\rm SFR}$ used by \citet{behroozi13a} in their work. \label{fig:sfrdens}}
\end{figure}

Figure~\ref{fig:ssfr} shows the distribution of specific star formation rates from my model. In the left, middle and right columns I show the distribution of specific 
star formation rate in cumulative bins of stellar mass (solid lines). The dashed lines show the results from the mock catalogs of \citet{watson2015}. 
The bottom panels show the results for centrals (blue) and satellites (red). I find that satellites in my model 
tend to be very quenched, possibly more so than in the SDSS group catalogs as described in \citet{watson2015}. Future extensions to the model 
based on the work of \citet{wetzel2013}, where satellite quenching is delayed, may be a useful in fixing this disagreement. 
Despite this difference the overall shift in the fraction of quenched galaxies follows the qualitative trends of the data. Note that the width and location of the quenched peak is directly 
determined by parameters in the model above. Also, my model adds negligible amount of scatter to the star forming sequence of galaxies. Thus the fact that the 
output star forming sequence has a scatter that is similar to the star forming peak in the SDSS indicates that the scatter in halo mass accretion rates 
contributions in a significantly to the scatter in star formation rates observed in the SDSS. 

Finally, Figure~\ref{fig:sfrdens} shows the density of star formation as a function of redshift, commonly called the Madau plot \citep{madau1998}. The points show 
direct measurements of the star formation rate density from my model. The line and grey band are the summarized results from data used by \citet{behroozi13a}. 
I find that my model successfully predicts the star formation rate density of the Universe. This agreement is expected since \citet{behroozi13a} fit the Madau plot 
as part of their analysis. 

\begin{figure*}
\includegraphics[width=\textwidth]{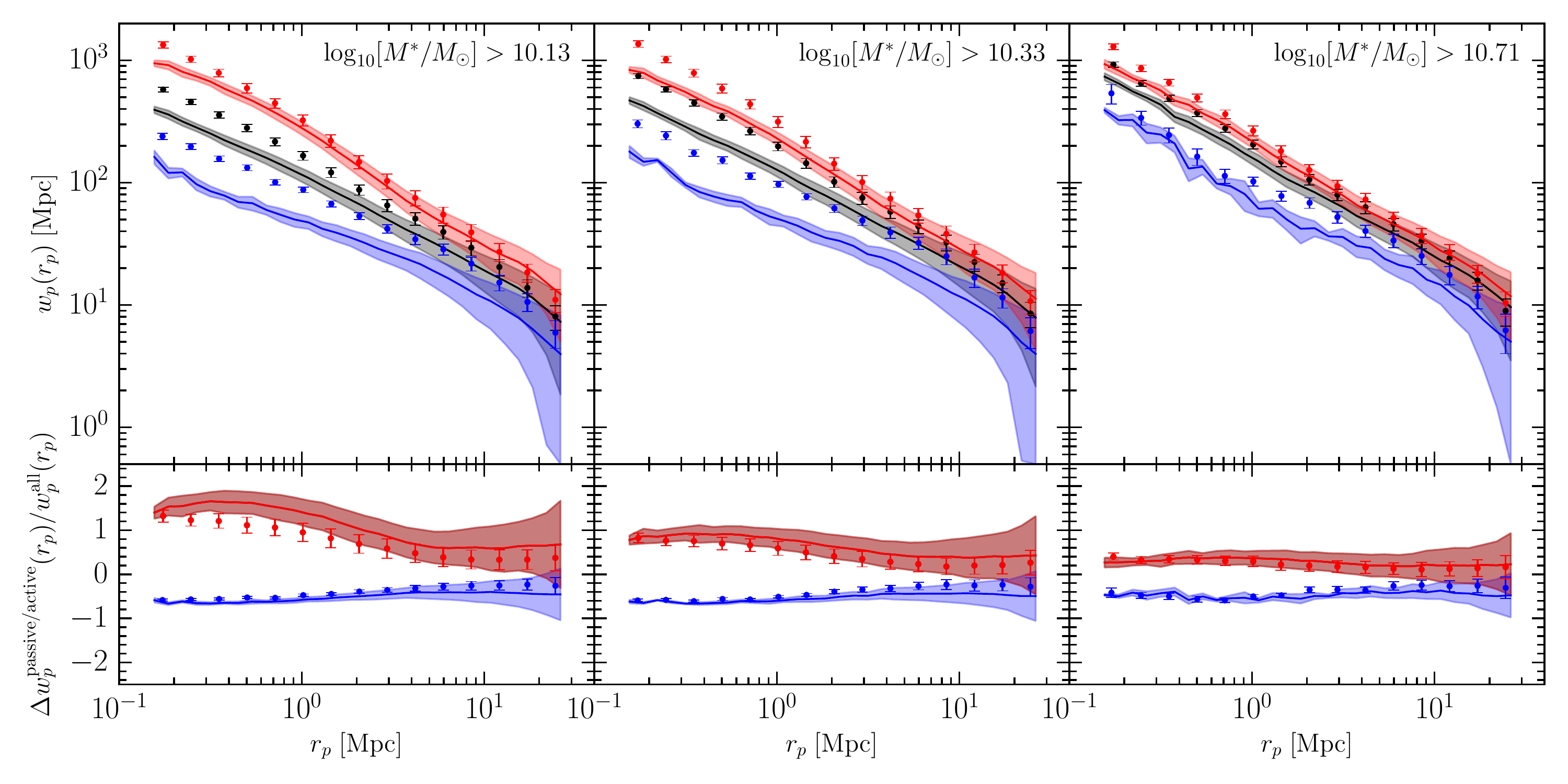}
\caption{The two-point clustering as a function of stellar mass and specific star formation rate. The points show the measurements from \citet{watson2015} 
of galaxies in the SDSS. The lines with bands show predictions from this work. Lines and points in red are quenched galaxies, in black are all galaxies, and in blue 
are active galaxies. the top panels show the projected correlation function and the bottom panels show the ratio of the correlation functions of active 
and passive galaxies to that of all galaxies. \label{fig:2pcf}}
\end{figure*}

\begin{figure*}
\includegraphics[width=\textwidth]{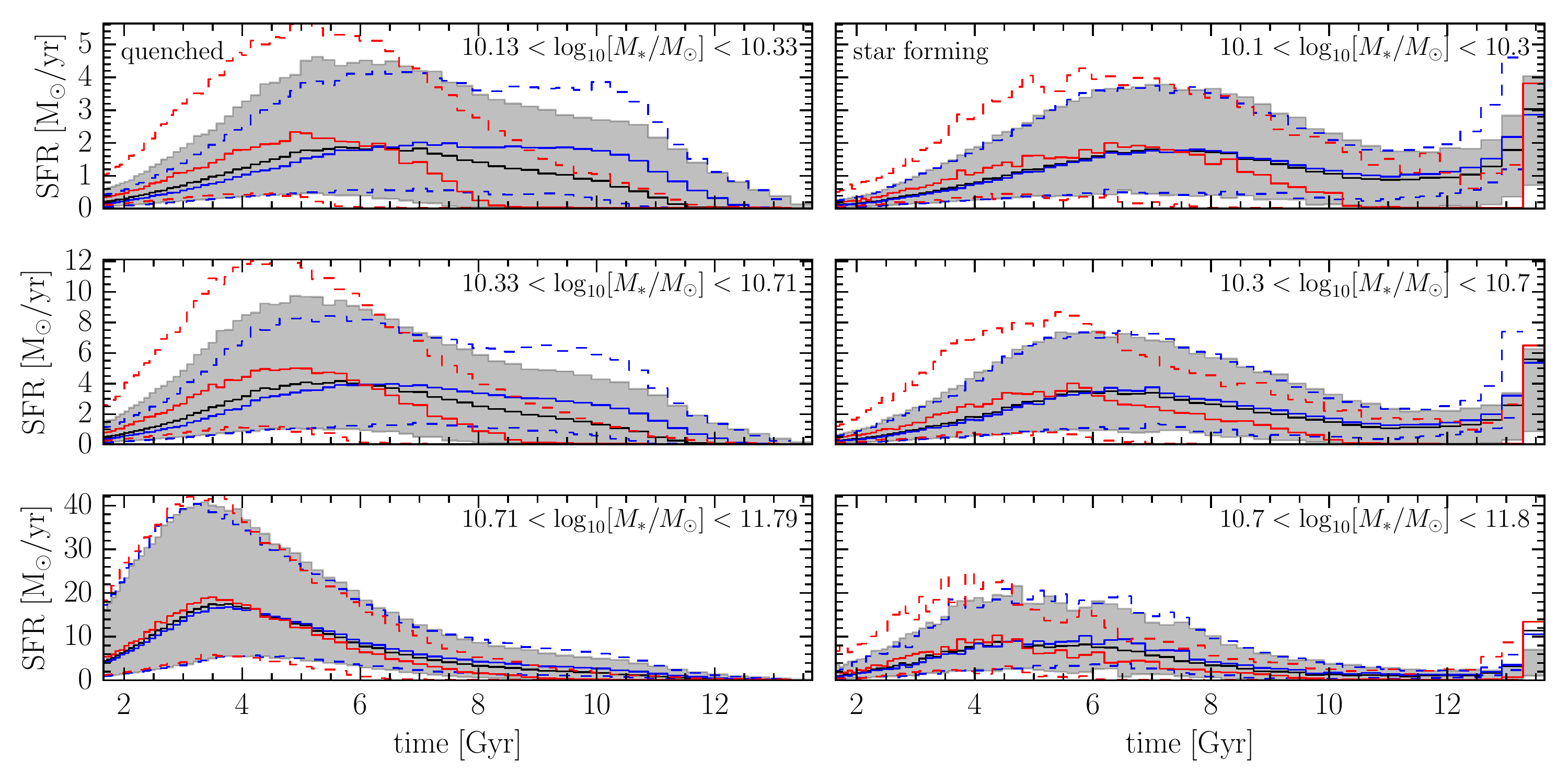}
\caption{Star formation histories along the most massive progenitor branches. The solid lines show the median star formation history 
for all galaxies (\textit{black}), central galaxies (\textit{blue}) and satellite galaxies (\textit{red}). The shaded grey band and dashed lines 
of the same color show the middle 50\% deviations about the median histories. \label{fig:sfhist}}
\end{figure*}

\begin{figure*}
\includegraphics[width=\textwidth]{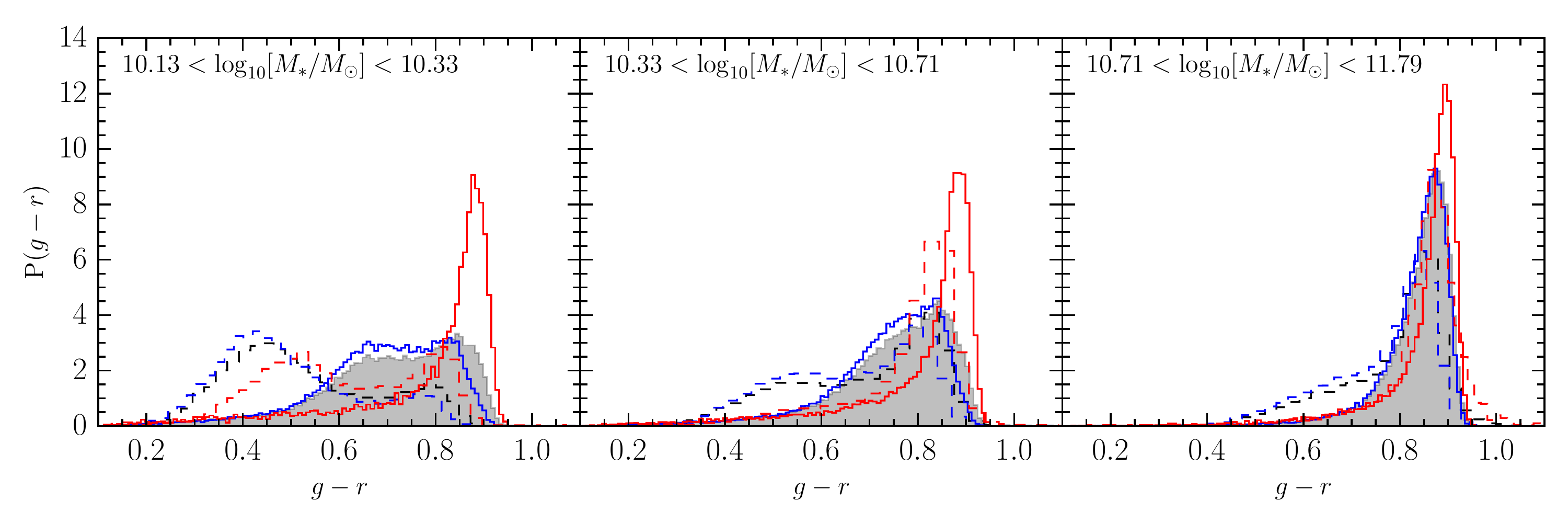}
\caption{SDSS $g-r$ color distributions for galaxies in stellar mass bins. All galaxies are shown in the grey shaded region, centrals are in blue and satellites are in red. 
The dashed lines of the same colors show the distributions in the mock catalog of \citet{watson2015} for all galaxies (\textit{black}), centrals (\textit{blue}) and satellites (\textit{red}). 
Note the $g-r$ color in the SDSS has been shifted down by 0.1 to account for an overall color difference between the colors in this work and the SDSS colors.
\label{fig:gmr}}
\end{figure*}

\subsection{Two-point Clustering}
Figure~\ref{fig:2pcf} shows the star formation rate and stellar mass dependent two-point clustering for my model compared to recent measurements in the 
SDSS \citep{watson2015}. The model error bars are from jackknife resampling of the simulation volume. I find that while the amplitude of the two-point clustering 
does not completely match the SDSS, the relative amplitude of the star forming 
and passive subsamples compared to the total clustering signals (shown in the bottom panels) does match the SDSS remarkably well. Note that \citet{hearin2015}
have shown that the mass accretion histories of halos are strongly influenced by their local environment through the tidal effects of other nearby halos. These effects 
generate the color dependent clustering in this model, directly through the mass accretion histories of individual halos in the merger trees. The 
partial success of this model indicates that assembly bias may play an important role in predicting the color dependent clustering seen in the SDSS, as proposed 
by \citet{hearin2013}.

\subsection{Stellar Mass Assembly Histories and Galaxy Colors}
Finally, Figures~\ref{fig:sfhist} and \ref{fig:gmr} show the stellar mass assembly histories and predicted $g-r$ colors of the galaxies in the SDSS bands. 
In Figure~\ref{fig:sfhist}, I show the star formation histories in bins of stellar mass (\textit{rows}) and for very quenched and very star forming galaxies 
(\textit{columns}). The solid black line and grey bands show the median and middle 50\% of the stellar mass assembly histories. The blue and red lines 
show the same data for centrals and satellites respectively. As expected, smaller galaxies tend to form the bulk of their stars later, whereas larger galaxies 
tend to form their stars quite early. This feature, noticed by \citet{conroy2009}, arises naturally from the combination of the \citet{behroozi13a} constraints 
and the $N$-body merger trees \citep{neistein2006}. Note also that the quenching of the star formation in massive halos is built directly in the \citet{behroozi13a} constraints. Finally, 
satellites tend to quench their star formation earlier than centrals, resulting in redder colors (see Figure~\ref{fig:gmr}) on average. 

Figure~\ref{fig:gmr} shows the SDSS $g-r$ colors of galaxies in cumulative stellar mass bins. I have used the \verb+FSPS+ code from \citet{conroy2009b,conroy2010} with
solar metallicity and a \citet{chabrier2003} initial mass function.\footnote{One must take care when running \texttt{FSPS} over the star formation histories. In particular, for the coarse star formation histories generated by the simulation merger tree, I found that accurate results could only be obtained by running the SPS assuming a piecewise constant star formation history with careful sub-integrations between each snapshot.} I find that my model produces consistently red satellites at all stellar masses. It also produces 
some amount of color bimodality, with the amount of color bimodality increasing as the stellar mass decreases. However, there are serious disagreements between my 
simple model and SDSS data as represented by the \citep{watson2015} mock catalog (dashed lines in the figure). (Note the $g-r$ color in the SDSS has been shifted down by 0.1 
to account for an overall color difference between my mock and the SDSS colors.) In particular, low stellar mass central galaxies 
appear to be too red. This feature is consistent with the distributions of sSFR's in Figure~\ref{fig:ssfr} where the star forming sequence as a whole appears to be a bit 
too quenched. Future work to directly fit the SMAD to the SDSS data should improve the model in these respects.  

\section{Conclusions}\label{sec:conc}
In this work, I have presented a general technique to model the galaxy-halo connect through the stellar mass assembly distribution (SMAD). 
In my implementation of this technique, the stellar mass assembly rates as constrained by \citet{behroozi13a} are used to construct an 
example SMAD. This model is then are integrated directly over halo merger trees from $N$-body simulations, accounting for mergers of 
dark matter halos. I find that my model qualitatively reproduces the bimodal clustering of the halos as a function of their star formation rates. In this model, 
this bimodality is due to the correlations in the mass accretion histories of halos with their large scale environments \citep{hearin2015}. 
My overall approach is very much like modern SAMs, but is directly statistical in nature. The SMAD parameterizes the ensemble of star formation rates 
at a fixed set of halo properties instead of attempting to predict the star formation rate correctly for any individual halo. Thus my model follows in the 
tradition of SHAM, CAM, HOD or CLF models which attempt to constrain the properties of galaxy formation directly from data as opposed to making 
\textit{a priori} predictions. 

Future work which uses the framework outlined here to fit for the parameters which govern 
$P(dM_{*}/dt|\mathbf{X},a)$ using the one- and two-point statistics of the observed galaxy population should yield interesting constraints on the 
growth of galaxies. Indeed, the general class of models discussed in this work can naturally reproduce observations of galactic conformity \citep[e.g.][]{weinmann2006,wang2010,robotham2013,kauffmann2013,phillips2014,knobel2015} 
and other potential signatures of assembly bias. These models also allow for natural extensions which can work to erase some of the assembly 
bias, such as adding correlated stochasticity in time to the stellar mass assembly rates. Simple extensions which, for example, randomly increase the 
star formation rates of galaxies in certain regimes, can account for other known effects like star forming BCGs in cooling flows, star bursts due to mergers, 
etc. Systematic changes in the stellar mass assembly rates of galaxies, say after they have reached their peak mass, may also be a useful modification \citep[e.g.][]{wang2007,wetzel2013}. 
This model combined with stellar population synthesis models can be used to directly predict the colors and/or spectra of galaxies in broad band filters, as demonstrated 
in this work. Additionally, this model will provide predictions for the rates of SNe as a function of environment, stellar mass and redshift, as well 
as the mean star formation rate density as a function of time or any other statistic which can predicted from an SPS code. These extensions and more 
detailed work fitting this model directly to data, such as clustering, lensing and redshift space distortions, will be the subject of future work. 

\section*{Acknowledgements}
I thank Andrew Hearin and Risa Wechsler for comments and encouragement during this work. I also than Peter Behroozi, Doug Watson 
and Frank van den Bosch for useful comments and discussion, and Yao-Yuan Mao for assistance with computing the 
projected correlation functions and simulation post-processing. This work made extensive use of
the NASA Astrophysics Data System and the \verb+arxiv.org+ preprint server.

\bibliographystyle{mn2e_good}
\bibliography{refs}

\label{lastpage}

\end{document}